\documentclass[usenatbib]{mnras}

\usepackage{newtxtext,newtxmath}

\usepackage[T1]{fontenc}

\DeclareRobustCommand{\VAN}[3]{#2}
\let\VANthebibliography\thebibliography
\def\thebibliography{\DeclareRobustCommand{\VAN}[3]{##3}\VANthebibliography}

\usepackage{graphicx}	
\usepackage{amsmath}	
\usepackage{hyperref}   
\usepackage{orcidlink}
\usepackage{natbib}
\usepackage{url}

\title[Lunar Farside RFI]{Modelling Radio Frequency Interference on the Lunar Farside}

\author[C. K. Ashe et al.]{
Charlie K. Ashe\orcidlink{0009-0009-3198-2768},$^{1}$
Ella J. Marshall\orcidlink{0009-0003-2523-1665},$^{2}$
Evan F. Keane\orcidlink{0000-0002-4553-655X},$^{1}$
Steve Prabu\orcidlink{0000-0003-3165-6785},$^{3}$
Richard Lynch,$^{4}$
\newauthor
David R. DeBoer\orcidlink{0000-0003-3197-2294}$^{3,5}$
\\
$^{1}$School of Physics, Trinity College Dublin, College Green, Dublin 2, D02 PN40, Ireland\\
$^{2}$School of Physics and Astronomy, The University of Edinburgh, Edinburgh, EH9 3JZ, UK\\
$^{3}$Sub-department of Astrophysics, University of Oxford, Oxford, OX1-3RH, UK\\
$^{4}$Heliospace, 2448 Sixth St, Berkeley, CA, 94710, USA\\
$^{5}$Radio Astronomy Laboratory, University of California, Berkeley, CA, 94720 USA
}

\date{Accepted XXX. Received YYY; in original form ZZZ}

\pubyear{\the\year{}}

\begin{document}
\label{firstpage}
\pagerange{\pageref{firstpage}--\pageref{lastpage}}
\maketitle

\begin{abstract}
Technological progress is a double-edged sword in the world of radio astronomy: improvements in antennas, receivers, and digital backends increase sensitivity, but those same advances, coupled with rapid growth in terrestrial and orbital communications, have resulted in a steep rise in radio-frequency interference (RFI). 
Modern datasets have become increasingly contaminated by anthropogenic emissions that are difficult to remove. These emissions can mimic or obscure astrophysical transients and narrow band signals, complicating the search for objects such fast radio bursts (FRBs), pulsars, and technosignatures. In light of these challenges, the lunar farside has recently been proposed as a site for radio-quiet observations, motivating missions such as the Lunar Farside Transients and Technology Telescope (LFT3), amongst many others. However, planned lunar and cislunar missions may introduce both intended and unintended radio emissions into this protected environment. Here we use a LFT3-like instrument to model the evolution of the apparent lunar farside RFI environment over the coming years. Using a catalogue of planned lunar satellites and models for their intended and unintended emissions, we estimate the received RFI over the $0.1$-$2700$~MHz frequency range. We find that while intended emissions are concentrated near communications bands, particularly around $2.4-2.6$~GHz, unintended emissions contaminate broad regions of the HF, VHF, and UHF bands. We further find that, under the assumptions of our model, uniform UEMR shielding of at least $30$~dB is required to place the maximum interference below the sensitivity of a LFT3-like instrument; more sensitive telescopes would correspondingly require higher UEMR shielding. Our calculations predict that the lunar farside will remain mostly usable for LFT3-like observations, but will become increasingly contaminated as the lunar satellite population grows. This highlights the decreasing window of opportunity for measurements of the lunar farside's radio quiet environment and should inform future mission design, timescales and policy choices for the protection of Moon-based radio science. 
 
\end{abstract}




\section{Introduction}
One of the biggest challenges in modern radio astronomy is radio frequency interference (RFI) from terrestrial and orbital technology. Radio telescopes are designed to detect weak natural emission from large distances, making them vulnerable in the presence of nearby, comparably stronger, radio emission created by human technology. This anthropogenic radiation can be intentional, such as broadcast radio, television, telecommunications, and satellite down-links, or unintentional, such as leakage from imperfectly-shielded electrical devices. RFI can both raise the effective noise floor across the spectrum (broadband interference, see e.g. \citealt{pkb+15}) and produce transient features that can mimic astrophysical signals (narrowband interference, see e.g. \citealt{swl16}). The latter is particularly obstructive in experiments targeting rare or narrowband astrophysical signals, including radio transients and technosignatures~\citep{daj+26}. RFI is now so prevalent that most observing frequencies are contaminated at least some of the time~\citep{rv24}.  

Protections for small fractional bands in the electromagnetic spectrum are in place in most jurisdictions, managed by the United Nations' agency the International Telecommunication Union (ITU\footnote{\url{www.itu.int}}) \citep{specm}. Most of the spectrum is utilised for transmissions. Radio astronomy is thus `blind' to many frequencies, e.g. the FM band. The challenges presented by RFI have only grown with time as the global radio environment has become populated with terrestrial transmitters and consumer electronics. A major tactic to minimise the deleterious effects of RFI has been to locate observatories in ever more remote locations; the currently-under-construction SKA-Low telescope in Western Australia is the exemplar~\citep{bbh+26}. This tactic seems to have reached the limit of its usefulness; in recent years large satellite constellations have been launched designed specifically to transmit powerful beams of radio emission to the most remote locations on Earth. The impacts of these satellites for ground-based astronomy are many, not only in the radio~\citep{dwb+23,gts+23,bdw+24,mhd+26} but also in the optical~\citep{tib+20,mop+22,khch23,hm24} and other bands~\citep{lsm23}.

Space-based radio observatories avoid \textit{some} terrestrial limitations, most importantly the opacity and variability of the Earth's ionosphere. However, observatories in orbit around the Earth or the Moon
are still subject to interference from the Sun, spacecraft, and anthropogenic transmitters (see e.g. \citealt{kgp+23}). For the purposes of radio astronomy, even moving a radio telescope to space does not necessarily provide a radio-quiet environment. 

The lunar \textit{farside} on the other hand is exceptional in respect to its radio quietness. Because the Moon is tidally locked to the Earth, its farside is permanently geometrically shielded from direct radio emission generated on or near the Earth. The absence of an ionosphere also enables observations at frequencies $\lesssim30$~MHz, a frequency regime that is at best very difficult, and often impossible, to study from Earth's surface~\citep{sbk+23}. At higher frequencies, the lunar farside also allows access to bands that are heavily contaminated on Earth. This combination of low RFI and broad spectral accessibility has motivated a wide range of lunar farside instruments. These include for instance Chang'E-4~\citep{Jia2018Change4}, Chang'E-6~\citep{Li2024Change6}, ROLSES-1~\citep{hbm+26}, LuSEE Night~\citep{bbd+23} and LFT3~\citep{daj+26}. However, the lunar farside will not remain radio quiet indefinitely. There are a growing number of lunar surface missions, communication relays, navigation systems, science payloads, and orbital infrastructure concepts being proposed for the coming decades. Spacecraft that transmit from lunar orbit, or operate on or near the lunar farside, will likely introduce intended or unintended emissions into an environment whose scientific value depends specifically on its radio quietness. It stands to reason that the emerging lunar radio environment must be assessed before large-scale activity alters it permanently. 

In this paper, we present a model for the evolving RFI environment on the lunar farside. In \S~\ref{sec:motivation} we give context for our work in more detail. The specifics of our modelling are detailed in \S~\ref{sec:model}. We present the results of our simulations, which project the RFI environment over a 5-year scale, in \S~\ref{sec:results}. In \S~\ref{sec:discussion} we discuss the implications of these results before concluding in \S~\ref{sec:conclusion}.



\section{Background}\label{sec:motivation}

The impact of RFI is particularly severe because radio astronomy is sensitivity limited by nature. For an ideal continuum observation, the minimum root-mean-square flux-density can be expressed through the radiometer equation~\citep{dtws85} as
\begin{equation}\label{eq:radiometer}
S_{\mathrm{min}} =
\frac{2 k_{\mathrm{B}} T_{\mathrm{sys}}}
{\eta_s A_{\mathrm{eff}}\sqrt{n_{\mathrm{pol}} \Delta \nu \tau}} \;,
\end{equation}
where \(T_{\mathrm{sys}}\) is the system temperature, \(A_{\mathrm{eff}}\) is the effective collecting area, \(\Delta \nu\) is the usable bandwidth, \(\tau\) is the integration time, \(n_{\mathrm{pol}}\) is the number of summed polarisations, and \(\eta_s\) represents the efficiency of the system. RFI degrades sensitivity in two complementary ways. Weak or unresolved interference can contribute additional effective noise, while strong interference requires the removal of affected samples or channels, reducing the available bandwidth and integration time~\citep{mrs22}. The sensitivity gain through increasingly large collecting areas and wide-band receivers can therefore be compromised by a radio-contaminated environment~\citep{bhg22}.

The Lunar Farside Transients and Technology Telescope, LFT3, is a proposed lunar farside radio instrument designed to take advantage of the radio-quiet environment of the Moon to conduct unique science. In this work, we use a LFT3-like instrument as the reference receiver for modelling the evolving lunar RFI environment. The assumed LFT3 system covers an observing range from $0.1$ to $2700$~MHz using three antenna components. These antennae are summarised in Table\ref{tab:LFT3_Antenna}. This wide frequency coverage makes LFT3 scientifically powerful, but also sensitive to a broad range of interference. Low-frequency observations are vulnerable to broadband and narrowband leakage from satellite electronics, while the upper part of the band overlaps with common spacecraft communications frequencies. 

However, advancements in space technology lead to a new risk: the pollution of the lunar RF environment by cislunar and lunar satellites. There are a large number of lunar missions currently being planned for launch over the next decade. Most missions will likely have unmeasured and unintended emissions typically in the low-frequency range, as well as intended emission in the so-called industrial, scientific and medical (ISM) $2.4-2.5$~GHz. With these satellites, we will see changes in the RF environment of the lunar farside. As of now, the lunar farside is the only location in the inner solar system that is RFI-free. Over the next decade, that will no longer be true. It is clear that some quantification of this RFI increase must be made, both for the sake of future missions and to highlight the importance of shielding for launched satellites and the timeline for LFT3 and other lunar scientific missions.

\section{Model}\label{sec:model}
\begin{figure*}
	\includegraphics[width=0.8\textwidth]{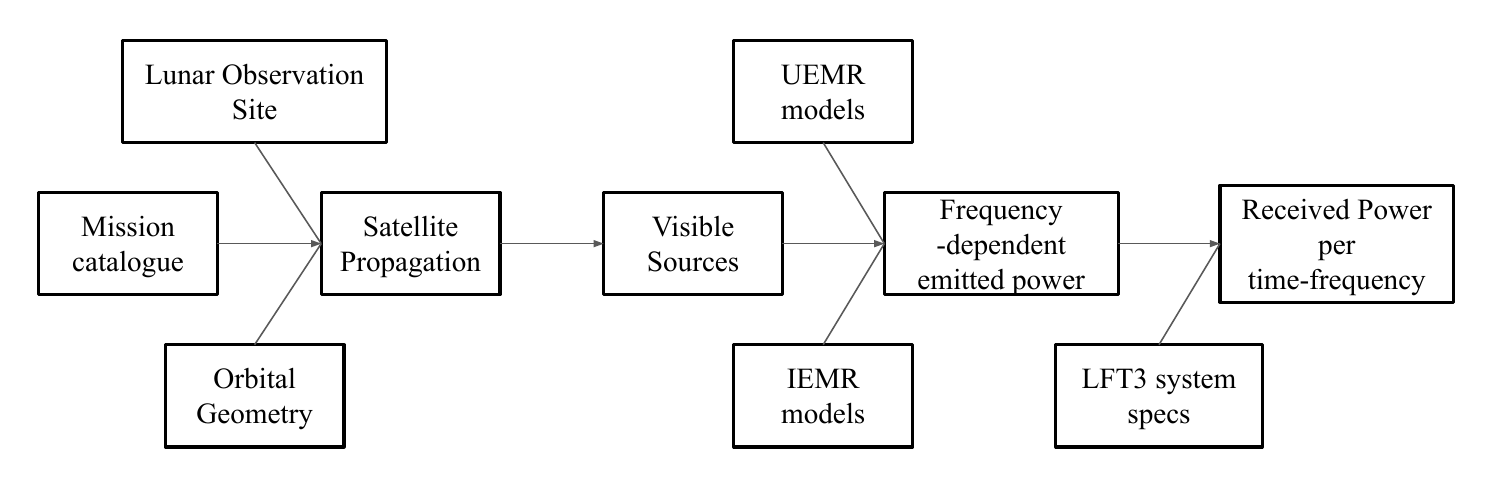}
    \caption{Overview of the model pipeline.}
    \label{fig:model_overview}
\end{figure*}

We model the evolving radio-frequency environment on the lunar farside using FEARLESS: the Farside Environment Analysis of RF Lunar-orbiter EmissionS Simulation. The code repository is open source, modular and new instruments can easily be added. We have in the first instance modelled the signals seen by LFT3-like receivers at the target LFT3 landing site (23.789°S, 182.137°E, \citealt{daj+26}) but an interested reader could obtain their own results for their chosen antenna specifications and chosen lunar farside location. We model RFI over time, from the increasing population of lunar and cislunar satellites. At each time step, the position of each satellite is propagated relative to a fixed observing site on the lunar farside. Spacecraft below the local horizon are discarded, while visible sources are retained with their corresponding range, azimuth, and elevation. For each visible source, the emitted power is evaluated as a function of frequency models for their intended electromagnetic radiation (IEMR) and unintended electromagnetic radiation (UEMR). The received power is computed by applying the frequency-dependent effective area and directional gain of the relevant LFT3 antenna. At each time step, the total power (Stokes $I$) contributions from all visible satellites are summed to produce the total RFI spectrum for that time step. We chose a time step of 500 s. This cadence was selected as a compromise between temporal resolution and computational cost, given the large number of frequency channels and the multi-year duration of the simulations. An overview of the modelling pipeline is given in Figure\ref{fig:model_overview}. 

\begin{table}
	\centering
	\caption{Characteristics of the antennas to be used in the LFT3 system. The gain is given in dBi, which is decibels with respect to an isotropic antenna that sees equally well across $4\pi$ steradians.}
	\begin{tabular}{lccccc} 
		\hline
		Antenna & Band & $\nu_{\rm min}$ & $\nu_{\rm max}$ & No. Beams & Peak Gain\\
        & & (MHz) & (MHz) & & (dBi) \\
		\hline
		Dipole & HF & 0.1 & 50.0 & 1 & 2.15\\
		Patch & VHF & 60.0 & 260.0 & 1 & 8.00\\
		Vivaldi Array & UHF & 300.0 & 2700.0 & 10 & 23.80\\
		\hline
	\end{tabular}
    \label{tab:LFT3_Antenna}
\end{table}

\subsection{LFT3 Instrument Assumptions}
The receiver model we use is based on the current mission concept for the LFT3 mission~\citep{daj+26}. LFT3 is assumed to observe over a broad frequency range, from $0.1 - 2700$~MHz, using three antenna systems: a low-frequency dipole, a VHF patch antenna, and a UHF Vivaldi array. The frequency ranges and peak gains of these three components are summarised in Table~\ref{tab:LFT3_Antenna}. 
For each antenna, the effective area is computed from the gain according to
\begin{equation}
    A_{\rm eff}(\nu) = \frac{\lambda^2 G(\nu)}{4\pi}, \label{eq:Aeff}
\end{equation}
where $\lambda$ is the observing wavelength and $G(\nu)$ is the antenna gain. 
The model uses the peak gain and beamweight models to determine the gain as a function of frequency and direction, allowing received power to depend on the apparent azimuth and elevation of each source. This distinction is important because a satellite with strong emissions may contribute weakly to the received power if it lies in a region of low antenna response. 

\subsection{Mission Catalogue}
The satellite population is defined by a mission catalogue containing the lunar and cislunar missions included in the simulation. For each mission, the catalogue specifies the orbital parameters required for propagation, as well as any available information on operational dates and transmitter properties. Only spacecraft with sufficient public information to assign approximate orbital and emission properties are included. Where exact parameters are unavailable, representative values have been adopted based on given details and comparable mission designs. These assumptions are recorded in the mission catalogue so that individual entries can be revised as more information becomes available. It is important to note that the parameters of these missions are likely to evolve with time and those chosen here may not reflect mission realities in the future. The full catalogue of satellites and their orbital parameters
is given in Table~\ref{tab:mission_catalogue}. 

\subsection{Emission Model}
Emission is evaluated on a frequency grid spanning $0.1-2700$~MHz with a resolution of $0.1$~MHz. This range covers the full LFT3 observing band. IEMR is transmitted deliberately by a satellite, whereas UEMR represents leakage from systems on the spacecraft. Distinction between IEMR and UEMR is necessary as each has a 
different physical origin and each is
modelled differently. 

IEMR components are assigned per-satellite using an input catalogue describing the emission type (narrowband or broadband), central frequency, equivalent isotropically radiated power (EIRP), and the bandwidth. In the current model, the IEMR catalogue is predominantly centred in frequency in the $2.4-2.6$~GHz range, with a spectrum that is well described as Gaussian. Each IEMR component is evaluated at its given frequency and converted from EIRP to received power at the antenna using the satellite-observer distance and the effective projected area of the relevant LFT3 antenna. Gaussian IEMR components are modelled as:
\begin{equation}
    P_{{\rm IEMR},i}(\nu) = P_{{\rm rec},i} \exp \left[-\frac{1}{2} \left(\frac{\nu - \nu_{0,i}}{\sigma_i} \right)^2 \right], \label{eq:iemr_gaussian}
\end{equation}
where $\nu_{0,i}$ is the central emission frequency, $\sigma_i$ is the spectral width, and $P_{\rm rec}$ is the received power at the emission peak. 

The UEMR catalogue contains a set of measured or representative emission bands. Each satellite's UEMR is described by minimum and maximum frequencies, 
minimum, maximum, and average EIRP values, an EIRP scatter, and an occurrence fraction. At each iteration a random sample is drawn to decide as to whether the UEMR for each satellite is `on' or `off' with the relative frequency decided by the occurrence fraction. When selected, the EIRP is determined by randomly sampling a Gaussian distribution according to the given mean of, and scatter in, the EIRP. The spectral shape is modelled differently depending on whether the signal is narrow- or broad-band; we arbitrarily define the distinction between these to be whether or not $f_{\rm max}-f_{\rm min}$ is less than or greater than $1$~MHz. For broadband UEMR the spectral shape is simply Gaussian according to the spectral parameters in the catalogue. 
Narrowband UEMR components are also Gaussian but with centre frequency drawn randomly from a uniform distribution between $f_{\rm min}$ and $f_{\rm max}$, and standard deviation subsequently chosen so that $95\%$ of the emission remains within the $[f_{\rm min},f_{\rm max}]$ range. In both cases, the profile is normalised so that the integrated power across the band matches the sampled EIRP.

Harmonics can be optionally added for both IEMR and UEMR. For a fundamental feature at frequency $\nu_0$, the $n^{\rm th}$ harmonic is placed at $n\nu_0$. Harmonic features are modelled using Gaussian profiles, with the $n$th harmonic assigned a width $n$ times larger than the width of the fundamental. The harmonic peak heights are estimated using the relative maxima of the sidelobes of a sinc-squared response. The total emitted spectrum for satellite i is then written as:
\begin{equation}
    P_{\rm emit,i}(\nu) = 
    P_{\rm IEMR,i}(\nu) +
    P_{\rm UEMR,i}(\nu) +
    P_{\rm harmonics,i}(\nu),
    \label{eq:emitted_power}
\end{equation}
where $P_{{\rm IEMR},i}(\nu)$ is the intended emission contribution, $P_{{\rm UEMR},i}(\nu)$ is the unintended emission contribution, and $P_{{\rm harm},i}(\nu)$ represents any harmonics included in the model. 


\subsection{Geometry and Received Power}
At each simulation timestep, the position of every satellite is propagated in a Moon-fixed coordination system. The observer is fixed at the planned LFT3 landing site on the lunar farside. For each satellite, the model computes the range to the observer and the apparent azimuth and elevation above the local horizon. Satellites below the horizon are assumed to be hidden by the Moon's body and do not contribute to the received power at that timestep. For a visible spacecraft, the flux $F$ (in watts per square metre) received at the observer is estimated from the EIRP (in watts) and the distance between the satellite and the observer $R$; for the i$^{\rm th}$ satellite this is:
\begin{equation}
    F_{\rm i}(\nu, t) = \frac{{\rm EIRP}_{\rm i}(\nu, t)}{4\pi R_{\rm i}(t)^2}\;.
\end{equation}
The corresponding received power is then calculated by multiplying this incident flux by the effective collecting area of the relevant antenna in the direction of the satellite,
\begin{equation}
    P_{\rm rec,i}(\nu, t) = 
    {\rm PFD_{ii}}(\nu, t) A_{\rm eff}(\nu, \theta_{\rm i}, \phi_{\rm i}),
\end{equation}
where $\theta_i$ and $\phi_i$ represent the apparent direction of the satellite in the local sky. For each iteration across the entire LFT3 frequency range the appropriate beam-pattern response is applied~\citep{daj+26}; to maximise gain beams are oriented towards local zenith.
Given the very large fractional bandwidth of LFT3 and having three different antenna types the beam correction varies significantly as a function of frequency. While satellites are rarely located at boresight, at low frequencies the primary beam is sufficiently broad that most remain within. As frequency increases they become more typically seen in side-lobes. Despite the off-axis detections, the satellite brightness is such that the detected signal levels are still quite significant; this is true across the entire LFT3 band.

\subsection{Satellite Shielding}
To investigate to what extent improved satellite design could reduce the contamination of the lunar radio environment, a model of satellite electromagnetic shielding is included. Shielding is applied only to the UEMR emitted by each satellite before propagation to the lunar observer. The shielding effectiveness is calculated as an attenuation $S_{\rm sh}$ in decibels. The shielded UEMR spectrum of satellite $i$ can be expressed as
\begin{equation}
P_{{\rm UEMR},i}^{\rm sh}(\nu,t) = 
P_{{\rm UEMR},i}(\nu,t) 10^{-S_{\rm sh}/10},
\end{equation}
where $P_{{\rm UEMR},i}(\nu,t)$ is the unshielded emitted UEMR power, and $P_{{\rm UEMR},i}^{\rm sh}(\nu,t)$ is the emitted power after shielding. 

We evaluate shielding effectiveness values of $S_{\rm sh}=0, 5, 10, 20, 30,$ and $40$~dB, but any range of values can be chosen in the model. A value of $0$~dB represents the baseline, unshielded model, while larger values represent progressively stronger levels of suppression. The same shielding effectiveness is applied to all modelled satellites in each simulation to isolate specifically the effect of satellite-level attenuation on the power of the UEMR received.

\section{Results}\label{sec:results}

The primary outputs of the simulation are satellite-visibility statistics and time-frequency received-power spectra. These products allow us to identify which parts of the LFT3 observing bands are most at-risk of contamination, how the RFI environment evolves as new satellites are launched, and whether the dominant interference is associated with particular regions of the sky. In this section, we first examine the changing number and sky distribution of visible satellites, then quantify the resulting unintended and intended electromagnetic radiation received across the LFT3 observing band.

\subsection{Satellite Visibility}
\begin{figure}
	\includegraphics[width=\columnwidth]{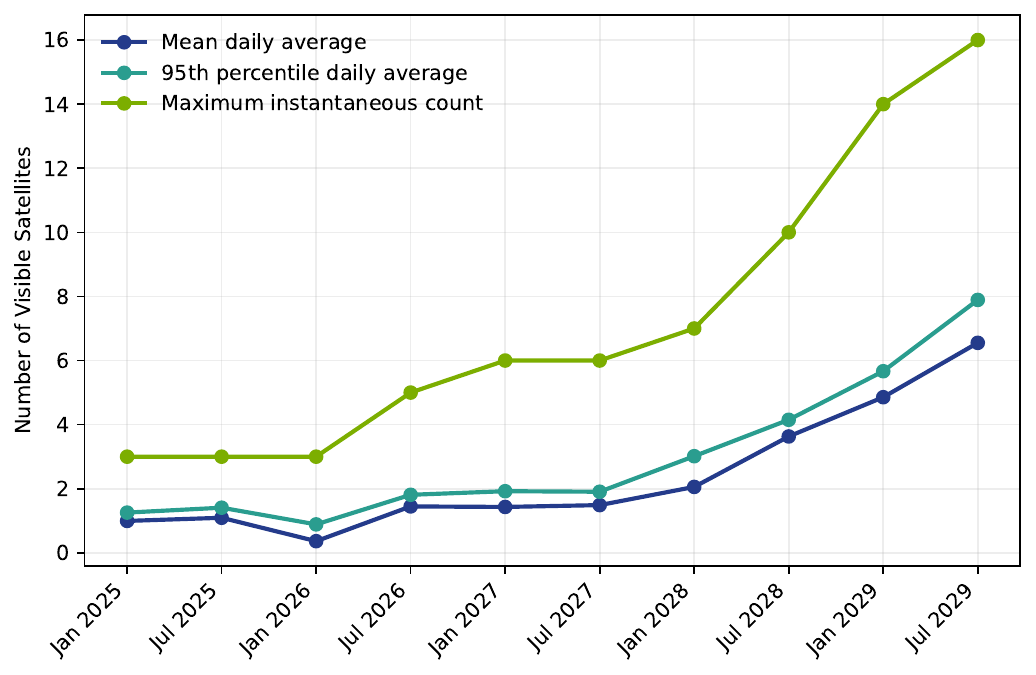}
    \caption{Summary statistics of visible-satellite counts for January and July from 2025 to 2029. The maximum represents the largest instantaneous visible-satellite count recorded during the month, the mean and 95th percentiles were computed from daily mean values within each month.}
    \label{fig:vis_satellites}
\end{figure}

Figure \ref{fig:vis_satellites} shows the simulated number of satellites visible from the LFT3 observing site in January and July from 2025 to 2029. A satellite is considered visible when it lies above the local horizon and is therefore not occulted by the Moon. The maximum value gives the largest instantaneous number of visible satellites recorded during each month, while the mean and 95th percentile are computed from daily mean visibility values. 

The simulation predicts a clear increase in the number of visible satellites over the modelled period. From 2025 to 2027, the mean daily number of visible satellites remains low, typically one to two. The maximum instantaneous count remains at six or below during this period, indicating that the lunar farside still experiences frequent intervals in which only a small number of satellites are visible above the horizon. From 2028 onwards, however, the visible population grows rapidly. By July 2028, the maximum count reaches ten, and by 2029 it rises to approximately sixteen. The mean daily visibility count also increases substantially, reaching more than six visible satellites by July 2029. 

This growth in the maximum numbers of satellites is important as RFI contamination is not characterised by the long-term average number of satellites alone. Short intervals in which many satellites are simultaneously visible can dominate the instantaneous RFI environment. The increasing separation between the mean, 95th percentile, and maximum curves indicates not only a higher average level of satellite visibility, but also a higher probability of short-duration, high-contamination intervals. In other words, the RFI environment becomes both brighter and more variable as the satellite population grows.

\begin{figure*}
	\includegraphics[width=0.75\textwidth, trim = 0mm 5mm 0mm 0mm, clip]{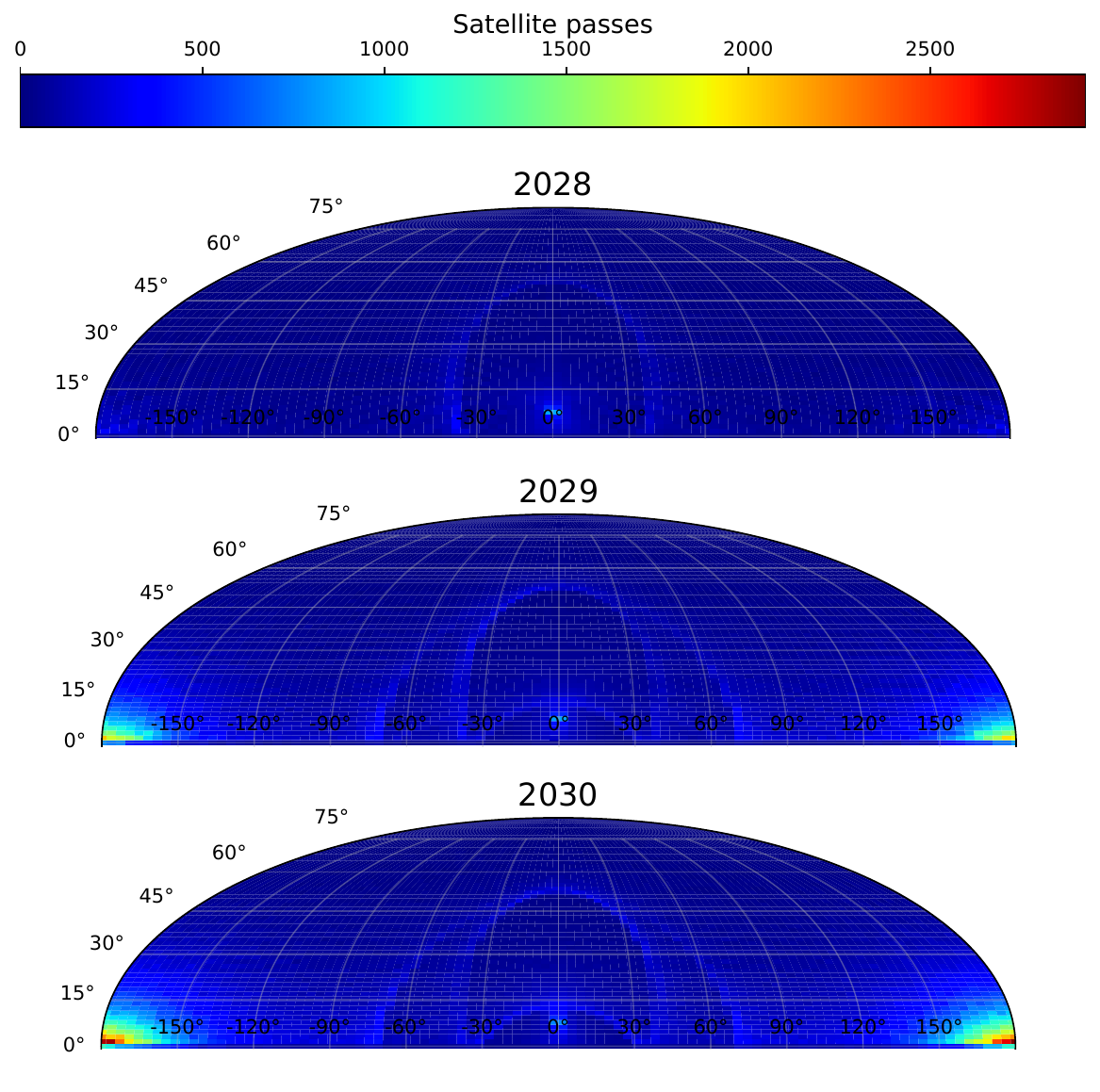}
    \caption{Sky-position maps of satellite-pass density for 2028, 2029, and 2030. Colour indicates the number of satellite passes through each azimuth-elevation bin, with dark blue corresponding to a low-pass density, and yellow to red corresponding to a high-pass density.}
    \label{fig:colourmap_satellites}
\end{figure*}

\subsection{Sky Distribution}
Figure \ref{fig:colourmap_satellites} shows sky-position maps of  satellite-pass density for 2028, 2029, and 2030. The maps are binned in azimuth and elevation, with colour scale indicating the number of satellites passes through each sky bin. These maps show that the growing satellite population does not appear uniformly across the local sky. Instead, the highest pass densities occur at low elevations, close to the local horizon, and with visible structure associated with the simulated orbital geometries.

The concentration of satellite passes becomes increasingly pronounced with time. In 2028, most areas of the sky remain at a relatively low pass density, although satellite orbital structure is already visible across the sky. By 2029, this structure is clearer and the number of passes increases across a larger fraction of the visible hemisphere. By 2030, the strongest concentrations occur at low elevations, particularly near the edges of the projected sky map. However the increase is not confined to a few `bad' locations, but instead affects a broad range of azimuths and elevations. This reduces the availability of consistently quiet skies, especially for low-frequency antennas with broad primary beams.

\subsection{Unintended and Intended Electromagnetic Radiation}

\begin{figure*}
	\includegraphics[width=0.9\textwidth]{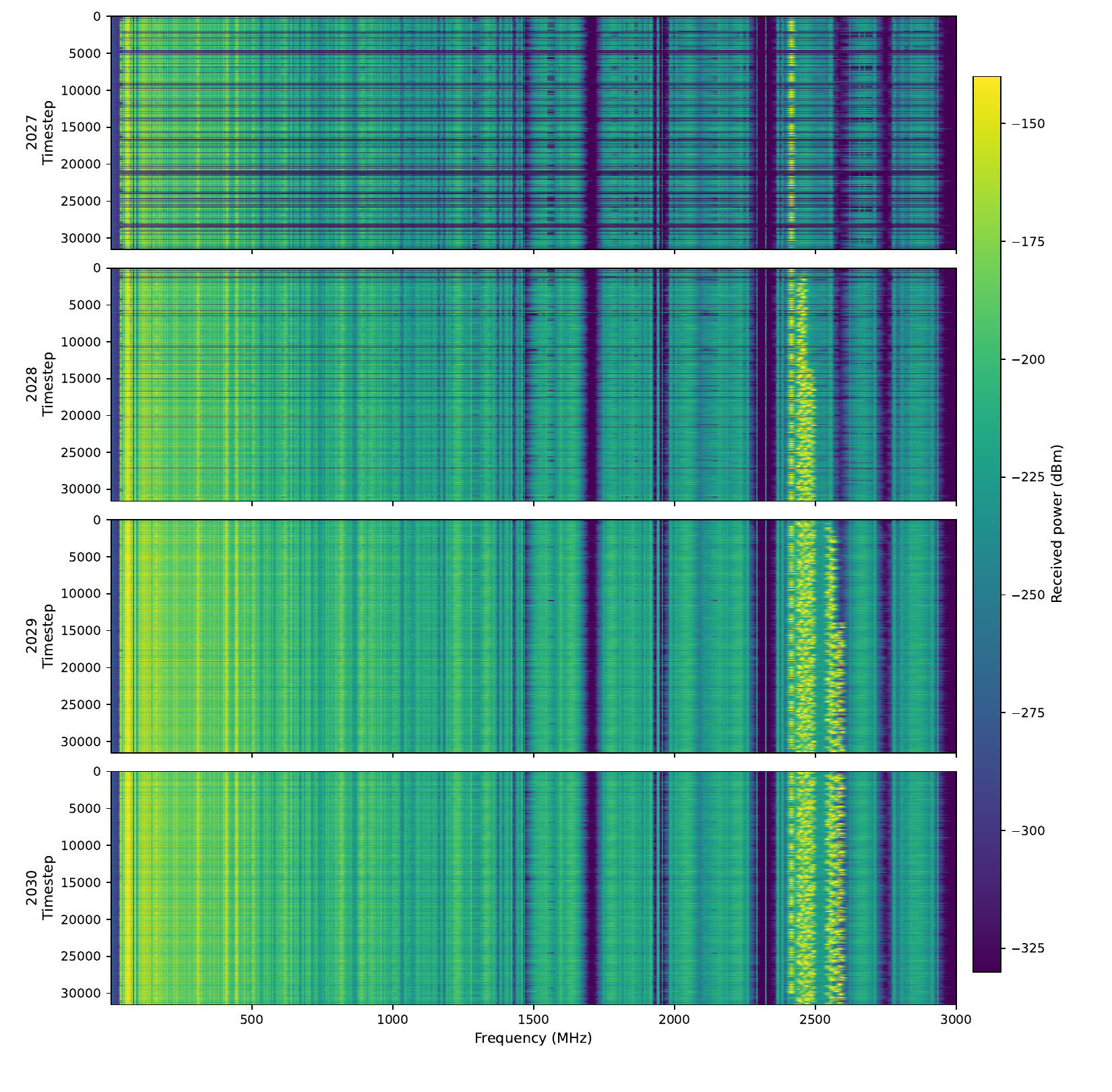}
    \caption{Simulated unintended electromagnetic radiation (UEMR) and intended electromagnetic radiation (IEMR) received by an LFT3-like instrument from 2027 to 2030. Each panel shows received power as a function of frequency and simulation timestep for one calendar year for the $0.1-3000$~MHz region.}
    \label{fig:waterfall}
\end{figure*}

Figure \ref{fig:waterfall} shows the simulated received power from 2027 to 2030. Each panel shows received power as a function of frequency and simulation timestep. The UEMR component forms a broad, nearly continuous background across much of the simulated frequency range, while the IEMR contribution appears as more spectrally localised features, prominent around $2.4-2.6$~GHz. 

The simulations show persistent contamination from unintended emissions even in 2027. The spectrum contains both narrowband and broadband features, with their apparent intensity varying over time as satellites move into and out of view and as their range from the observing site changes. By 2028, the density of these features increases noticably. In 2029 and 2030, the spectra become densely occupied, with many frequencies showing persistent contamination.

This follows naturally from the increasing number of visible satellites shown in Figure \ref{fig:vis_satellites}. As more satellites are present above the horizon, more leakage sources contribute to the received spectrum. 

At higher frequencies, the UEMR background remains continuous but generally weaker, with several relatively low-power spectral regions visible. Superimposed on this UEMR background are the localised IEMR features, which appear primarily in assigned communications bands. These features are weak or absent in 2027, but become more prominent and occupy a larger fraction of the band by 2029 and 2030 as additional satellites enter the modelled population. By the end of the simulated period, the IEMR component is persistent and contains multiple narrowband features at different frequencies.

We emphasise that the upper-frequency behaviour of the modelled UEMR component is determined by the available emission catalogue. We discuss this further in \S~\ref{sec:discussion}.

\begin{figure*}
	\includegraphics[width=0.8\textwidth]{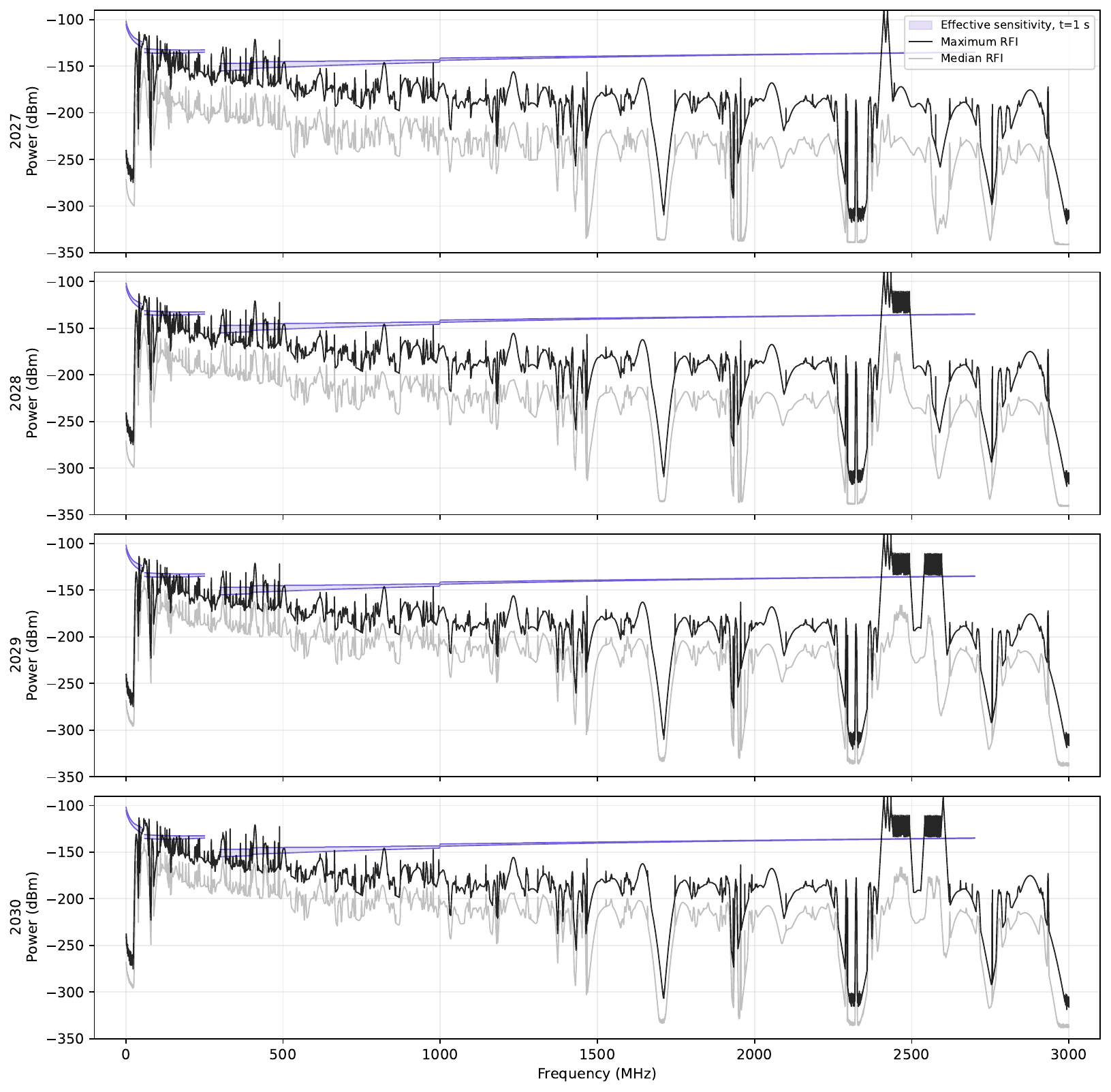}
    \caption{Maximum and median simulated RFI spectrum plotted against the effective sensitivity of a LFT3-like instrument for a $1$-s integration. The black curve shows the maximum received RFI power as a function of frequency for each year, the grey curve shows the median received RFI power as a function of frequency for each year, and the purple curve shows the effective sensitivity at each frequency.}
    \label{fig:med_spectra}
\end{figure*}

Figure \ref{fig:med_spectra} compares the median and maximum simulated RFI spectrum with the effective sensitivity of a LFT3-like instrument for $1$-s integration between 2027 and 2030. This comparison is not intended to provide a detectability calculation. Instead, it provides a useful indication of which parts of the band are most likely to be affected and how the RFI spectrum changes over time. Across the lower-frequency region, the median RFI spectrum contains many features that approach the effective sensitivity curve, especially as the satellite population increases. Consistent with the UEMR time-frequency spectra, the strongest low-frequency contamination occurs below $1500$~MHz. At higher frequencies, the median RFI remains low over much of the band, but distinct features appear around the intended-emission bands near $2.4-2.6$~GHz. These features become more prominent and contaminate a larger effective bandwidth from 2028 onward. The maximum shows how short-lived, geometrically-aligned events can create spectra that exceed the median by tens of decibels across the band. Below $1500$~MHz and between $2.4-2.6$~GHz especially, these maxima lie frequently above the sensitivity curve. Even frequencies with modest median contamination can experience strong interference intermittently. The maxima in the intended-emissions bands increase in both amplitude and bandwidth with each year. The growing separation between the maximum and median each year also highlights how the variability of the environment increases with time. 

\subsection{Effects of Satellite Shielding}
\begin{figure}
	\includegraphics[width=\columnwidth]{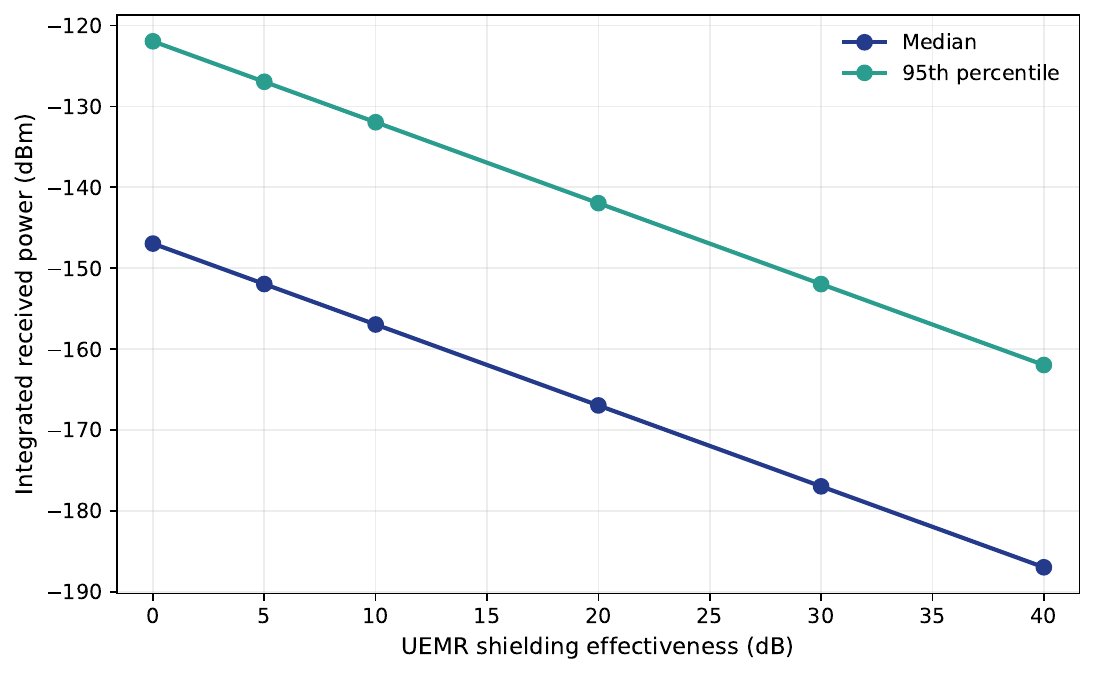}
    \caption{Integrated received UEMR power as a function of satellite shielding effectiveness. The dark curve shows the median power, and the lighter curve shows the 95th percentile.}
    \label{fig:shield_sens}
\end{figure}

Figure~\ref{fig:shield_sens} shows the integrated UEMR power received as a function of the effectiveness of putative satellite shielding. The integrated received power decreases as the shielding increases. In the unshielded case, the median integrated received power is approximately $-147$~dBm, while the 95th-percentile value is approximately $-122$~dBm. As the shielding effectiveness increases, both the median and 95th-percentile powers decrease by 
the applied attenuation, within noise fluctuations. For example, increasing the shielding effectiveness from $0$ to $40$~dB reduces the median integrated received UEMR power from approximately $-147$~dBm to $-187$~dBm, corresponding to a factor of $10^4$, or a $99.99\%$ reduction in linear received power. 

\begin{figure*}
	\includegraphics[width=0.8\textwidth]{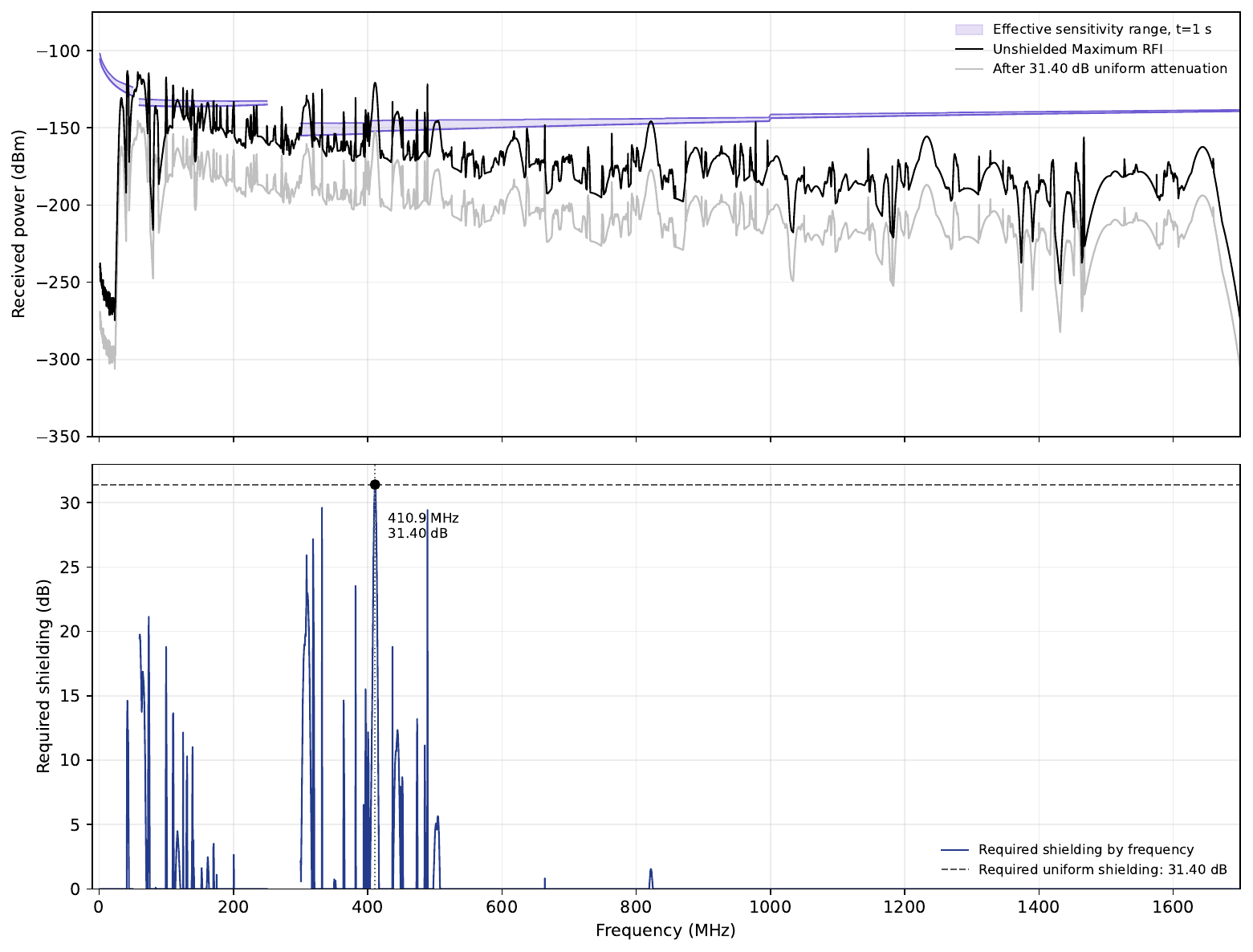}
    \caption{Frequency-dependent shielding requirements for a LFT3-like instrument. The top panel compares the unshielded maximum UEMR spectrum with the spectrum after applying $31.40$~dB of attentuation. The shaded region shows the sensitivity of a LFT3-like instrument for a $1$~s integration. The bottom panel shows the minimum shielding required for each frequency.}
    \label{fig:shielding_required}
\end{figure*}

Figure~\ref{fig:shielding_required} shows the maximum UEMR spectrum without shielding, the spectrum after application of a uniform attenuation, and the effective LFT3 sensitivity range for an integration of $1$~s. The lower panel shows the attenuation required independently at each frequency. The most contaminated frequency is at $410.9$~MHz. A uniform shielding effectiveness of $31.40$~dB is required to place the maximum UEMR below the LFT3 sensitivity threshold at every analysed frequency. This attenuation corresponds to a reduction in emitted UEMR power by a factor of approximately $120$. 

\section{Discussion}\label{sec:discussion}
The simulations presented here indicate that the lunar farside radio environment could evolve rapidly over the next several years. In early periods of the modelled timeframe, the LFT3 site retains many intervals with few visible satellites, and the corresponding RFI environment is comparatively sparse. By 2028 to 2029, however, both the number of visible satellites and the density of received emission increase substantially. The result is a transition from an environment that is contaminated intermittently to an environment in which some level of satellite-associated RFI is present near-constantly. 

Importantly, the most problematic component is not necessarily the strongest intended transmitter. Intended emissions near $2.4-2.6$~GHz are clearly visible in the simulations, but they are spectrally localised. Although \textit{this} emission still represents a loss of usable bandwidth, the necessity for communications relays, e.g. retrieving astronomical data on Earth, means some loss of this nature is almost inevitable in the absence of long-distant fibre-optic cabling across the lunar surface~\citep{pzrd26}. In contrast, the unintended emissions are distributed across a much wider frequency range and are present throughout the HF, VHF, and UHF portions of the LFT3 band. This makes UEMR more difficult to mitigate with simple filtering. If the leakage model used here is representative of future lunar missions, then unintentional emissions may become the dominant limitation on the preservation of the lunar farside as a radio-quiet site. 

The current model is limited by the frequency coverage of the UEMR catalogue. Directly measured UEMR is included only up to approximately ~$230$~MHz; at higher frequencies, the model is populated using harmonics of the catalogued emission, calculated up to harmonic order n = 8. While this provides a representative leakage spectrum and avoids leaving large, unmodelled gaps, it cannot be interpreted as an evidence-based description of satellite UEMR at these frequencies. Leakage above $230$~MHz is poorly characterised in available public data. There is limited public information on the frequencies at which the onboard electronics of proposed lunar missions generate unintended emission. Much of the available evidence comes from low-frequency observations \citep{dwb+23,bdw+24,StLkRFI}, which imposes an artificial boundary on the UEMR that can be modelled with confidence. The apparent reduction in UEMR at higher frequencies therefore should not be interpreted as evidence that satellite leakage is intrinsically weak or absent there, but as evidence that improvements to the model are necessary to accurately quantify leakage at these frequencies. 


This would require a dedicated measurement campaign targeting spacecraft and satellite UEMR over a wider frequency range. A series of measurements with facilities such as CHIME with its outrigger stations \citep{CHIMEOutriggers2025} would be invaluable for constraining leakage in the $400-800$~MHz range. We intend to conduct such a campaign with the Allen Telescope Array, availing of its high tracking rate ability \citep{Welch_2009}. Such measurements would benefit radio astronomy more generally, and would allow our model to become ever more realistic by replacing at-present \textit{representative} parts of our catalogue with accurate measurements of satellite leakage. Without these measurements, the impact of RFI above several hundred MHz remains uncertain.

The spatial distribution of satellite passes suggests that some possible mitigation strategies may be possible through observation strategy and antenna beam response. For example, avoiding low-elevation directions could reduce exposure to the highest-density regions of the satellite sky. However, this is not a complete solution. The simulated sky maps show that, by 2030, satellite passes occur across a wide range of sky positions, with most bins having at least a hundred passes a month. For low-frequency antennas with broad beams covering a large fraction of a hemisphere, and for science cases simply requiring large sky coverage, avoiding contaminated directions will be difficult and limiting.

Operational coordination with satellite operators could reduce the impact of intended emissions, for example, by scheduling quiet periods or restricting transmissions during sensitive observations, but this approach would be ineffective for UEMR unless the satellites are designed, tested, and regulated with radio astronomy in mind. Improved shielding, stricter limits on out-of-band leakage, and pre-launch measurements of satellite emission spectra would directly reduce the level of contamination seen at lower frequencies. None of these protective measures are intractable and, given the current unique pristine conditions on the lunar farside, the necessary steps for preservation are relatively minor. 

The results, following the implementation of shielding, help quantify the scale of mitigation that would be required for the LFT3 system to observe optimally in this environment. Figure~\ref{fig:shield_sens} shows that that increasing satellite shielding produces a direct reduction in both the typical and high-percentile integrated UEMR power received. Figure~\ref{fig:shielding_required} shows how the shielding requirement can be determined by the strongest individual spectral feature. The most contaminated frequencies require at least $31.40$~dB of attenuation. Applying this value uniformly is sufficient enough to place the maximum UEMR below the sensitivity threshold of LFT3 across the analysed frequency range. But this result is specific to the sensitivity and bandwidth assumed for an LFT3-like instrument. It should not be interpreted as a universal shielding standard for lunar satellites. Many radio astronomy instruments are currently in development for lunar farside deployment. More sensitive instruments will detect lower received powers, and would therefore require greater suppression of satellite UEMR to satisfy the same criterion. Similarly, longer integrations would require better shielding as the effective sensitivity of a radio telescope to weak signals improves with integration time as per Equation~\ref{eq:radiometer}.

Several limitations should be kept in mind when interpreting the results of the simulation. The mission catalogue is incomplete and contains representative orbital and operational parameters where final values are unavailable. Furthermore, additional architectures continue to be proposed that are not included in this simulation. For example, \citet{KoblickCasey2026} recently evaluated five- and six-satellite tulip-orbits for lunar south-pole position, navigation, timing and relay constellations. The UEMR model is also  uncertain. The exact received powers and affected frequencies may change as better emission data become available as missions come closer to deployment. This also affects the shielding requirements, as UEMR may in reality be better or worse depending on the reality of the distribution of shielding effectiveness in the satellite population. It is one of the primary goals of LFT3 to measure UEMR in situ. 

The qualitative trend is robust: increasing the number of active lunar and cislunar satellites increases the probability that one or more emitters are visible from the lunar farside at any given time. Our initial analyses presented here suggest it is very likely that a large number of satellite passes will be visible to radio astronomy facilities each day. Unless future satellites are designed with radio astronomy requirements in mind, the farside is likely to become progressively less radio-quiet, deteriorating from its currently pristine state. This has direct implications not only for the timing of LFT3, but other, similar missions as well. The scientific value of the lunar farside is greatest before the growth of satellite populations produce persistent RFI contamination. There is a limited window in which the farside can be observed under conditions close to its natural radio quietness.  


\section{Conclusions}\label{sec:conclusion}
We have presented FEARLESS, a simulation framework for modelling the evolving radio-frequency environment of the lunar farside. Using the mission parameters of a LFT3-like instrument, as well as a catalogue of planned lunar and cislunar satellites, we estimated the RFI produced by intended and unintended satellite emissions over the next several years. As a result of these simulations, we have been able to conclude the following about the RFI environment of the lunar farside.

\begin{itemize}
    \item The number of satellites visible from a LFT3-like farside site increases substantially over the modelled period. From 2025 to 2027, the mean number of visible satellites remains low, but by 2028 the mean and maximum visible counts rise rapidly, increasing the likelihood of intervals in which multiple satellites are contributing to the RFI environment. Satellite visibility is not uniformly distributed across the sky. The highest pass densities occur at low elevations, close to the local horizon. However, by 2030, satellite passes occur across a broad range of azimuths and elevations, reducing the availability of consistently quiet sky regions.

    \item Unintended electromagnetic radiation is the most challenging aspect of RFI in current simulations. While intended emissions remain concentrated near $2.4-2.6$~GHz, allowing for filtering or removal of contaminated bands, UEMR contaminates a much broader frequency range, including the HF, VHF, and lower-UHF portions of the LFT3 band. This makes UEMR difficult to mitigate through simple excision or avoidance. The apparent decline of UEMR about $1.5$~GHz is a limitation of available spectral data and not indicative of an absence of leakage at these frequencies. Dedicated measurements of satellite emissions across a wider frequency range are necessary to improve future forecasts of the lunar RFI environment. 

    \item Satellite shielding provides a direct means of reducing contamination from unintended emissions. Across the tested shielding values, the median and 95th-percentile integrated received UEMR powers decrease approximately in proportion to the applied attenuation. A comparison of different shielding levels with the effective sensitivity of a LFT3-like instrument shows that at least $31.40$~dB of uniform attenuation is required to place the maximum UEMR received below the sensitivity threshold. 

    \item Mitigation of RFI will require more than observing strategy alone. Avoiding low-elevation directions, or coordinating intended transmissions could reduce some contamination, but preserving the lunar farside as a radio-quiet environment will require satellite-level shielding, stricter leakage limits, and pre-launch emission measurements. The shielding requirement derived here is specific to a LFT3-like system. More sensitive future instruments (e.g. FARSIDE \citet{FARSIDE}) would require lower satellite emission levels. 

    \item The lunar farside remains an exceptional site for radio astronomy, but its radio quietness cannot be assumed to persist indefinitely. The growth of lunar and cislunar infrastructure creates a challenge for low-frequency science that is time-sensitive. While missions such as LFT3 can exploit a unique observing environment, the long-term scientific value of the farside will depend on early coordination between mission designers and the radio astronomy community.

\end{itemize}

\section*{Acknowledgements}
C. K. Ashe thanks the School of Physics in Trinity College Dublin, and the Institute of Physics for support via the Jocelyn Bell Graduate Scholarship Fund. This article is based on work supported by Breakthrough Listen.
Breakthrough Listen is managed by the Breakthrough Initiatives,
sponsored by the Breakthrough Prize Foundation.

\section*{CRediT Author Statement}
\textbf{Charlie K. Ashe}: Conceptualisation, Methodology, Software, Formal Analysis, Writing --- Original Draft, Writing --- Review \& Editing, Visualisation

\noindent
\textbf{Ella J. Marshall}: Conceptualisation, Methodology, Software, Data Curation, Formal Analysis, Writing --- Review \& Editing

\noindent
\textbf{Evan F. Keane}: Conceptualisation, Methodology, Supervision, Writing --- Original Draft, Writing --- Review \& Editing

\noindent
\textbf{Steve Prabu}: Conceptualisation, Methodology, Supervision

\noindent
\textbf{Richard Lynch}: Conceptualisation, Investigation

\noindent
\textbf{David R. DeBoer}: Conceptualisation, Software, Validation, Resources, Supervision

\section*{Code \& Data Availability}
The software developed for this work is called FEARLESS. We provide this open source\footnote{\texttt{https://github.com/ScienceMoonshot/Lunar-RFI-Study}} and encourage its use and development by others. 



\bibliographystyle{mnras}
\bibliography{example} 




\appendix

\section{}
\begin{table*}
    \centering
    \scriptsize
    \caption{Mission catalogue used in the FEARLESS simulation.
    The columns specify the mission name, semi-major axis $a$,
    eccentricity $e$, inclination $i$, argument of periapsis $\omega$,
    periapsis precession rate $\dot{\omega}$, right ascension of the
    ascending node $\Omega$, true anomaly $f$, start year, start month,
    and operational lifetime.
    [1] = \protect\citet{NASA_CAPSTONE};
    [2] = \protect\citet{KPLO2023};
    [3] = \protect\citet{Queqiao22025};
    [4] = \protect\citet{KHON12022};
    [5] = \protect\citet{LunarPathfinder2023};
    [6] = \protect\citet{Volta};
    [7] = \protect\citet{FireflyElytraDark};
    [8] = \protect\citet{AlpineLupine};
    [9] = \protect\citet{Luna26};
    [10] = \protect\citet{Chandrayaan4};
    [11] = \protect\citet{Andromeda};
    [12] = \protect\citet{Chen_2020DSL};
    [13] = \protect\citet{LUGO}}
    
    \label{tab:mission_catalogue}
    \resizebox{\textwidth}{!}{%
    \begin{tabular}{lccccccccccc}
        \hline
        Label & $a$ (km) & $e$ & $i$ (deg) & $\omega$ (deg) & $\dot{\omega}$ (deg d$^{-1}$) & $\Omega$ (deg) & $f$ (deg) & Year & Month & Lifetime (months) & Ref.\\
        \hline
        CAPSTONE & 37544.10 & 0.91 & 90.00 & 105.00 & -- & -- & -- & 2022 & 6 & 43 & 1\\
        KPLO & 1837.40 & 0.01 & 90.21 & 0.00 & -- & -- & -- & 2022 & 9 & 40 & 2\\
        Queqiao-2 & 9837.40 & 0.803 & 62.40 & 270.00 & 0.00 & 0.00 & -- & 2024 & 4 & 96 & 3\\
        Khon1 & 2230.00 & 0.00 & 62.00 & 90.00 & -- & 190.00 & -- & 2025 & 3 & 96 & 4\\
        Lunar Pathfinder & 5740.00 & 0.58 & 54.856 & 86.322 & -- & 0.00 & -- & 2026 & 1 & 60 & 5\\
        Volta1 & 1787.00 & 0.00 & 331.00 & 0.00 & -- & 270.00 & 0.00 & 2026 & 1 & 60 & 6 \\
        Volta2 & 1787.00 & 0.00 & 331.00 & 0.00 & -- & 270.00 & 120.00 & 2026 & 1 & 60 & 6 \\
        Volta3 & 1787.00 & 0.00 & 331.00 & 0.00 & -- & 270.00 & 240.00 & 2026 & 1 & 60 & 6 \\
        Elytra Dark \#1 & 1787.40 & 0.00 & 265.00 & 0.00 & -- & 185.00 & -- & 2026 & 2 & 72 & 7 \\
        Elytra Dark \#2 & 1787.40 & 0.00 & 265.00 & 0.00 & -- & 185.00 & -- & 2026 & 2 & 72 & 7 \\
        Alpine & 6500.00 & 0.00 & 90.00 & 0.00 & -- & 35.00 & 0.00 & 2026 & 6 & 72 & 8\\
        Lupine & 6500.00 & 0.00 & 90.00 & 0.00 & -- & 35.00 & 180.00 & 2026 & 6 & 72 & 8\\
        Luna26 & 1807.40 & 0.01 & 90.00 & 0.00 & -- & 35.00 & -- & 2028 & 1 & 12 & 9\\
        Chandrayaan-4 & 1837.40 & 0.00 & 108.00 & 0.00 & -- & 345.00 & -- & 2028 & 1 & 60 & 10\\
        Andromeda 1 & 6142.40 & 0.60 & 57.00 & 270.00 & -- & 0.00 & 0.00 & 2028 & 1 & 60 & 11\\
        Andromeda 2 & 6142.40 & 0.60 & 57.00 & 270.00 & -- & 0.00 & 60.00 & 2028 & 1 & 60 & 11 \\
        Andromeda 3 & 6142.40 & 0.60 & 57.00 & 270.00 & -- & 0.00 & 120.00 & 2028 & 1 & 60 & 11 \\
        Andromeda 4 & 6142.40 & 0.60 & 57.00 & 270.00 & -- & 0.00 & 180.00 & 2028 & 1 & 60 & 11 \\
        Andromeda 5 & 6142.40 & 0.60 & 57.00 & 270.00 & -- & 0.00 & 240.00 & 2028 & 1 & 60 & 11 \\
        Andromeda 6 & 6142.40 & 0.60 & 57.00 & 270.00 & -- & 0.00 & 300.00 & 2028 & 1 & 60 & 11\\
        Andromeda 7 & 6142.40 & 0.60 & 57.00 & 270.00 & -- & 90.00 & 0.00 & 2028 & 6 & 60 & 11 \\
        Andromeda 8 & 6142.40 & 0.60 & 57.00 & 270.00 & -- & 90.00 & 60.00 & 2028 & 6 & 60 & 11 \\
        Andromeda 9 & 6142.40 & 0.60 & 57.00 & 270.00 & -- & 90.00 & 120.00 & 2028 & 6 & 60 & 11 \\
        Andromeda 10 & 6142.40 & 0.60 & 57.00 & 270.00 & -- & 90.00 & 180.00 & 2028 & 6 & 60 & 11 \\
        Andromeda 11 & 6142.40 & 0.60 & 57.00 & 270.00 & -- & 90.00 & 240.00 & 2028 & 6 & 60 & 11 \\
        Andromeda 12 & 6142.40 & 0.60 & 57.00 & 270.00 & -- & 90.00 & 300.00 & 2028 & 6 & 60 & 11 \\
        Luna26 End & 1877.00 & 0.01 & 90.00 & 0.00 & -- & 35.00 & -- & 2029 & 1 & 12 & 9\\
        DSL 1 & 2037.40 & 0.00 & 30.00 & 0.00 & -- & 330.00 & 0.00 & 2029 & 1 & 60 & 12\\
        DSL 2 & 2037.40 & 0.00 & 30.00 & 0.00 & -- & 330.00 & 40.00 & 2029 & 1 & 60 & 12 \\
        DSL 3 & 2037.40 & 0.00 & 30.00 & 0.00 & -- & 330.00 & 80.00 & 2029 & 1 & 60 & 12 \\
        DSL 4 & 2037.40 & 0.00 & 30.00 & 0.00 & -- & 330.00 & 120.00 & 2029 & 1 & 60 & 12 \\
        DSL 5 & 2037.40 & 0.00 & 30.00 & 0.00 & -- & 330.00 & 160.00 & 2029 & 1 & 60 & 12 \\
        DSL 6 & 2037.40 & 0.00 & 30.00 & 0.00 & -- & 330.00 & 200.00 & 2029 & 1 & 60 & 12 \\
        DSL 7 & 2037.40 & 0.00 & 30.00 & 0.00 & -- & 330.00 & 240.00 & 2029 & 1 & 60 & 12 \\
        DSL 8 & 2037.40 & 0.00 & 30.00 & 0.00 & -- & 330.00 & 280.00 & 2029 & 1 & 60 & 12 \\
        DSL 9 & 2037.40 & 0.00 & 30.00 & 0.00 & -- & 330.00 & 320.00 & 2029 & 1 & 60 & 12 \\
        Andromeda 13 & 6142.40 & 0.60 & 57.00 & 270.00 & -- & 180.00 & 0.00 & 2029 & 1 & 60 & 12 \\
        Andromeda 14 & 6142.40 & 0.60 & 57.00 & 270.00 & -- & 180.00 & 60.00 & 2029 & 1 & 60 & 12 \\
        Andromeda 15 & 6142.40 & 0.60 & 57.00 & 270.00 & -- & 180.00 & 120.00 & 2029 & 1 & 60 & 12 \\
        Andromeda 16 & 6142.40 & 0.60 & 57.00 & 270.00 & -- & 180.00 & 180.00 & 2029 & 1 & 60 & 12 \\
        Andromeda 17 & 6142.40 & 0.60 & 57.00 & 270.00 & -- & 180.00 & 240.00 & 2029 & 1 & 60 & 12 \\
        Andromeda 18 & 6142.40 & 0.60 & 57.00 & 270.00 & -- & 180.00 & 300.00 & 2029 & 1 & 60 & 12 \\
        Andromeda 19 & 6142.40 & 0.60 & 57.00 & 270.00 & -- & 270.00 & 0.00 & 2029 & 6 & 60 & 12 \\
        Andromeda 20 & 6142.40 & 0.60 & 57.00 & 270.00 & -- & 270.00 & 60.00 & 2029 & 6 & 60 & 12 \\
        Andromeda 21 & 6142.40 & 0.60 & 57.00 & 270.00 & -- & 270.00 & 120.00 & 2029 & 6 & 60 & 12 \\
        Andromeda 22 & 6142.40 & 0.60 & 57.00 & 270.00 & -- & 270.00 & 180.00 & 2029 & 6 & 60 & 12 \\
        Andromeda 23 & 6142.40 & 0.60 & 57.00 & 270.00 & -- & 270.00 & 240.00 & 2029 & 6 & 60 & 12 \\
        Andromeda 24 & 6142.40 & 0.60 & 57.00 & 270.00 & -- & 270.00 & 300.00 & 2029 & 6 & 60 & 12 \\
        LUGO & 1801.92 & 0.02 & 86.00 & 180.00 & -- & 27.00 & -- & 2030 & 1 & 12 & 13 \\
        \hline
    \end{tabular}%
    }
\end{table*}


\bsp	
\label{lastpage}
\end{document}